\documentclass[journal]{IEEEtran}

\usepackage{url}
\usepackage{graphicx}
\usepackage{amsmath}
\usepackage{amsfonts}
\usepackage{amssymb, color}
\usepackage{algorithm}
\usepackage[noend]{algpseudocode}

\newtheorem{theorem}{Theorem}

\newtheorem{corollary}{Corollary}

\begin{document}

\title{Wireless Powered Communication Networks with Non-Ideal Circuit Power Consumption }

\author{Slavche Pejoski, Zoran Hadzi-Velkov, Trung Q. Duong, and Caijun Zhong

\vspace{-3mm}

\thanks{S. Pejoski and Z. Hadzi-Velkov are with the Faculty of Electrical Engineering and Information Technologies, Ss. Cyril and Methodius University, 1000 Skopje, Macedonia (email: slavchep@feit.ukim.edu.mk, zoranhv@feit.ukim.edu.mk).}
\thanks{T. Q. Duong is with the School of Electronics, Electrical Engineering and Computer Science, Queen's University Belfast, Belfast BT7 1NN, U.K. (email: trung.q.duong@qub.ac.uk).}
\thanks{C. Zhong is with the Institute of Information and Communication Engineering, Zhejiang University, Hangzhou 310027, China (e-mail: caijunzhong@zju.edu.cn).}

}

\markboth{}{Shell \MakeLowercase{\textit{et al.}}: Bare Demo of IEEEtran.cls for Journals} 

\maketitle

\begin{abstract}
Assuming non-ideal circuit power consumption at the energy harvesting (EH) nodes, we propose two practical protocols that optimize the performance of the harvest-then-transmit wireless powered communication networks (WPCNs) under two different objectives: (1) proportional fair (PF) resource allocation, and (2) sum rate maximization. These objectives lead to optimal allocations for the transmit power by the base station (BS), which broadcasts RF radiation over the downlink, and optimal durations of the EH phase and the uplink information transmission phases within the dynamic time-division multiple access (TDMA) frame. Compared to the {\it max-sum-rate} protocol, the {\it PF} protocol attains a higher level of system fairness at the expense of the sum rate degradation. The {\it PF} protocol is advantageous over the {\it max-sum-rate} protocol in terms of system fairness regardless of the circuit power consumption, whereas the uplink sum rates of both protocols converge when this power consumption increases.

\end{abstract}


\vspace{-3mm}

\section{Introduction}
The energy harvesting (EH) communication systems scavenge energy either from natural (such as, the solar or wind) [\ref{litnova1}], or man-made sources (such as, the radio-frequency (RF) radiation from power beacons) [\ref{litnova3}]. If the EH system relies on natural sources, it may be difficult to devise causal protocols that achieve sufficiently high rates, due to the intermittent nature of these energy sources. If an EH system is powered by RF radiation (e.g., a {\it wireless powered communication network} (WPCN)), it can even achieve rates comparable to that of conventional (non-EH) networks when intelligent policies for power and rate control are applied [\ref{lit2}]. Typically, the WPCNs are optimized for their spectral efficiency, i.e., their sum rate over the uplink is maximized by dynamic adaptation of the base station (BS) transmit power and durations of the EH phase and the information transmission (IT) phase in each time-division multiple access (TDMA) frame [\ref{lit1a}]. The WPCNs may achieve even higher sum rates if the EH users (EHUs) employ non-orthogonal multiple access (NOMA) transmissions over the uplink by spending only a part of the energy available in their batteries, while saving the rest of it for future use [\ref{lit19}]. However, despite its high spectral efficiency, a resource allocation strategy based upon the sum-rate maximization may unfairly distribute the system resources among the EHUs at different distances from the BS. Instead, an opportunistic scheduling policy may facilitate system fairness, such as, the {\it proportional fair (PF) scheduling} [\ref{lit6}], [\ref{litnova6}]. We here focus on maximizing the uplink sum rate in a proportionally fair way.

Most of the studies assume that the EHUs spend their harvested energy only for RF transmissions, thus significantly overestimating the predicted rates. In fact, the non-ideal electric circuitry of practical wireless transmitters consume significant additional power (e.g., AC/DC converter, analog RF amplifier, and processor), and need to be taken into account [\ref{lit1o}]-[\ref{litnova5}]. In this paper, we develop two optimal protocols for the TDMA-based WPCNs with non-ideal circuit power consumption;  The first one guarantees proportionally fair resource sharing among the EHUs, and the second ones maximizes the sum rate over the uplink of the WPCN.

\vspace{-3mm}
\section{System Model}
We consider a WPCN with a single BS and $K$ EHUs that operate in a wireless fading environment. Each network node is equipped with a single antenna. The network utilizes TDMA, and the time is divided into $M$ TDMA frames of equal duration $T$ (also referred to as the transmission epochs). Each epoch is divided into $K+1$ phases, an EH phase and $K$ successive IT phases, whose durations can be dynamically adjusted in each epoch. In epoch $i$, the duration of the EH phase, and the IT phase of $k$th EHU is $\tau_0(i)T$ and $\tau_k(i)T$, respectively, such that $\tau_{0}(i) + \tau_1(i) + \dots + \tau_K(i) = 1$. During the EH phase of epoch $i$, the BS broadcasts RF energy with an output power $p_{0}(i)$, which satisfies a maximum power constraint $P_{max}$ (i.e., $ 0 \leq p_{0}(i) \leq P_{max}$), and an average power constraint $P_{avg}$ (i.e., $ E[p_{0}(i)\tau_{0}(i)] \leq P_{avg}$).

The random channel follows a quasi-static block fading model, where each fading block coincides with a single epoch. The channel between the BS and $k$th EHU is assumed reciprocal, and its fading power gain in epoch $i$ is denoted by $x_{k}'(i)$. For convenience, we normalize these power gains as $x_{k}(i) = x'_{k}(i)/N_0$, where $N_0$ is the additive white Gaussian noise (AWGN) power. The average value of $x_{k}(i)$ is $\Omega_k = E[x'_{k}(i)]/N_0$, where $E[\cdot]$ denotes expectation. We assume perfect synchronization among the network nodes. The channel state information requirements are the following: the BS should know all $K$ fading links, $\{x_{k}(i)\}_{k = 1}^K$, whereas $k$th EHU should know only its own fading channel, $x_{k}(i)$.

\vspace{-1mm}
\subsection{Power Consumption Model for the Battery}
The EHUs are equipped with rechargeable batteries that have low energy storage capacity and high discharge rate. Specifically, when an EHU transmits information, it completely spends all of the harvested energy in its battery during the previous BS broadcast in that same epoch, which is typically referred to as the {\it harvest-then-transmit} strategy [\ref{lit2}], [\ref{lit1a}]. For the power consumption by the EHU, we apply a realistic model that incorporates both the transmit power and the non-ideal circuit power. Therefore, the total power consumed by the $k$th EHU in epoch $i$ is given by \cite{lit1o}
\begin{equation} \label{powermodel}
p_{T,k}(i) = \begin{cases}
P_k(i)+p_c, & P_k(i) > 0\\
0, & P_k(i)=0
\end{cases}
\end{equation}
where $P_k(i)$ is the transmit power of $k$th EHU, and $p_c$ is the non-ideal circuit power consumed by that EHU during its IT phase. Note that $p_c$ is fixed and independent of the EHU.

\section{Optimal resource allocation}
\label{sec1}
In epoch $i$, the amount of harvested energy by $k$th EHU during the EH phase is given by
\begin{equation} \label{Ek}
E_k(i) = \eta_k x_{k}(i) N_0 p_{0} (i)\tau_{0}(i) T,
\end{equation}
where $\eta_k$ is the conversion efficiency of $k$th EHU. Given the power consumption model (\ref{powermodel}), $P_k(i)$ is determined by
\begin{equation}
P_{k}(i) = \max\left\{0,\frac{E_k(i)}{\tau_{k}(i)T}-p_c\right\},
\end{equation}
where $E_k(i)$ is given by (\ref{Ek}). Therefore, the achievable rate of $k$th EHU in epoch $i$ is given by
\begin{equation} \label{Rk1}
r_{k}(i) = \tau_{k}(i) \log_2(1 + P_{k}(i) x_{k}(i)),
\end{equation}
whereas the average achievable rate over $M$ epochs is
\begin{equation} \label{Rk2}
\bar{R}_k=\lim_{M\rightarrow \infty}\frac{1}{M}\sum_{i=1}^M r_k(i).
\end{equation}

\vspace{-4mm}
\subsection{Proportional fair resource allocation }
\vspace{-0mm}
The first protocol, referred to as the {\it PF} protocol, maximizes the sum of the logarithmic rates achieved by the network users, $\sum_{k = 1}^K \log \bar R_k$ (c.f. [\ref{litnova6}], and references therein). Thus, when $M \to \infty$, we need to determine the optimal durations of the EH and IT phases and the optimal BS transmit power in each epoch by solving the following optimization problem:
\begin{eqnarray} \label{OP2}
\underset{\tau_{k}(i),\tau_{0}(i),p_{0}(i)} {\text{Maximize}} \ \sum_{k=1}^K\log\left( \frac{1}{M}\sum_{i=1}^M \tau_{k}(i)\log_2(1+x_{k}(i)P_{k}(i))  \right) \notag
\end{eqnarray}
\text{s.t.}
\vspace{-5mm}
\begin{eqnarray}
\begin{array}{ll}
&C1: P_{k}(i)  = \max\left\{0,\frac{E_k(i)}{\tau_{k}(i)T}-p_c\right\}, \forall i, \,\, 1\leq k \leq K \\ \notag
&C2: \frac{1}{M}\sum_{i=1}^M p_{0}(i) \tau_{0}(i) \leq P_{avg} \\ \notag
&C3: 0 \leq p_{0}(i) \leq P_{max}, \forall i \\ \notag
&C4: \sum_{k=1}^K \tau_{k}(i) = 1-\tau_{0}(i), \forall i \\ \notag
&C5: 0 < \tau_{k}(i) < 1, \forall i, \,\, 0 \leq k \leq K.
\end{array}
\end{eqnarray}
\vspace{-7mm}
\begin{equation}
\end{equation}

The solution of (\ref{OP2}) is given by the following theorem.

\begin{theorem}\label{teorem1}
The optimal transmit power of the BS is
\begin{equation} \label{sol11}
p_{0}^*(i) =
\begin{cases}
P_{max}, & \sum_{k=1}^K \frac{1}{\bar{R}_k}\frac{a_{k}(i)}{1-p_cx_{k}(i)+z_{k}(i)}> \lambda \\
0, & \text{otherwise}.
\end{cases}
\end{equation}
The optimal durations of EH and IT phases are determined by
\begin{equation} \label{sol21}
\tau_{0}^*(i) =\frac{1}{1+\sum_{k=1}^K\frac{a_{k}(i)P_{max}}{z_{k}(i)}}, \qquad \qquad
\end{equation}
\begin{equation} \label{sol31}
\tau_{k}^*(i) = \frac{a_{k}(i)P_{max}}{z_{k}(i)}\tau_{0}^*(i), \quad 1\leq k \leq K,
\end{equation}
respectively, where $a_{k}(i)=\eta_k N_0x_{k}^2(i)$. In (\ref{sol11}), (\ref{sol21}) and (\ref{sol31}), $z_{k}(i)$ are auxiliary variables that are determined by
\begin{eqnarray}
\label{sol41}\notag
z_{k}(i) = -(1-p_c x_{k}(i)) \qquad \qquad \qquad \qquad \qquad \qquad \\
\times \left[1+\frac{1}{W(-(1-p_c x_{k}(i))\, e^{-1-\beta_i P_{max} \bar{R}_k})}\right],
\end{eqnarray}
where $W(\cdot)$ is Lambert-$W$ function, and $\beta_i$ is found as the root of the following transcendental equation,

\begin{eqnarray}
\label{sol51}\notag
\sum_{k=1}^K\hspace{-3mm}&\frac{1}{\bar{R}_k}&\hspace{-3mm}\frac{a_{k}(i)}{1-p_cx_{k}(i)} W\left(-(1-p_c x_{k}(i))\,e^{-1-\beta_i P_{max}\bar{R}_k} \right)\\
&+&\hspace{-3mm}\beta_i+\lambda= 0
\end{eqnarray}
The constant $\lambda$ is determined from $C2$ in (\ref{OP2}) set to equality.
\end{theorem}

\begin{IEEEproof}
Please refer to Appendix A.
\end{IEEEproof}

In practice, the values of $\lambda$ and $\bar{R}_k$ may not be available in advance. For an online estimation of $\lambda$, we apply the stochastic gradient descent method [\ref{lit3b}], as
\begin{equation} \label{lamb_calc}
\hat{\lambda}(i) = \hat{\lambda}(i-1)+\gamma_0\left(\frac{1}{i-1}\sum_{n=1}^{i-1}p_{0}(n)\tau_{0}(n) - P_{avg}\right),
\end{equation}
where $\gamma_0$ is some small step size. The rate $\bar R_k$ can be also updated online according to a simple iterative rule,
\begin{equation} \label{Rk_calc}
\hat R_k(i) = \frac{i-1}{i} \, \hat R_k(i-1) + \frac{1}{i} \, r_k(i).
\end{equation}
Actually, each iteration of $\hat R_k(i)$ is based upon an ever increasing window size, equal to the elapsed session time $i$, which guarantees the maximization of $\sum_{k = 1}^K \log \bar R_k$ [\ref{lit6}, Lemma 4]. The practical implementation of the proposed policy at the BS is outlined by Algorithm 1.

\begin{algorithm}
{
\caption{PF protocol implementation at the BS} \label{alg1}
\begin{algorithmic}[1]
\State $\text{Initialize}$ $\hat \lambda$, $\hat R_k\ \forall k$. $\text{Set}$ $\gamma_0$ $\text{and}$ $i = 1$;
\Repeat
\State $\text{Determine}$ $b(i)=\sum_{k=1}^K \frac{1}{\hat {R}_k}\frac{a_{k}(i)}{1-p_cx_{k}(i)+z_{k}(i)}$;
\If {$b(i) > \hat \lambda$}
\State $\text{Calculate}$ $\tau_{0}(i)$ $\text{from}$ (\ref{sol21});
\State $\text{Calculate}$ $\tau_{k}(i)$ $\text{from}$ (\ref{sol31}), $\text{and feedback to EHUs}$;
\State $\text{Calculate}$ $r_k(i)$ $\text{from}$ (\ref{Rk1}), $\text{and feedback to EHUs}$;
\State $\text{Broadcast RF energy at}$ $p_0(i) = P_{max}$ $\text{for}$ $\tau_{0}(i) T$;
\Else
\State $\text{Set}$ $p_0(i) = 0$, $\tau_{k}(i) = 0$  $\text{and}$ $r_k=0,\ \forall k$;
\EndIf
\State $\hat \lambda \leftarrow \hat \lambda + \gamma_0 \left( \frac{1}{i} \sum_{n=1}^{i} p_0(n) \tau_{0}(n) - P_{avg} \right)$;
\State $\hat R_k \leftarrow \frac{i - 1}{i} \, \hat R_k + \frac{1}{i} \, r_k(i), \, \forall k$;
\State $i \leftarrow i + 1$.
\Until {$i \leq M$}
\end{algorithmic} }
\end{algorithm}

\vspace{-3mm}
\subsection{Sum rate maximization}
The second protocol, referred to as the {\it max-sum-rate} protocol, aims at maximizing the achievable average rate in the uplink of the WPCN, $\sum_{k=1}^K \bar R_k$, yielding
\begin{eqnarray} \label{OP1}
\underset{\tau_{k}(i),\tau_{0}(i),p_{0}(i)} {\text{Maximize}} \ \frac{1}{M}\sum_{i=1}^M\sum_{k=1}^K \tau_{k}(i)\log_2(1+x_{k}(i)P_{k}(i)) \notag
\end{eqnarray}
\text{s.t.}
\vspace{-5mm}
\begin{eqnarray}
\begin{array}{ll}
&C1: P_{k}(i)  = \max\left\{0,\frac{E_k(i)}{\tau_{k}(i) T} -p_c\right\}, \forall i, \, 1 \leq k \leq K \\ \notag
&C2: \frac{1}{M}\sum_{i=1}^M p_{0}(i) \tau_{0}(i) \leq P_{avg} \\ \notag
&C3: 0 \leq p_{0}(i) \leq P_{max}, \forall i \\ \notag
&C4: \sum_{k=1}^K \tau_{k}(i) = 1-\tau_{0}(i), \forall i \\ \notag
&C5: 0 < \tau_{k}(i) < 1, \forall i, \quad 0 \leq k \leq K.
\end{array}
\end{eqnarray}
\vspace{-11mm}
\begin{equation}
\end{equation}

The solution of (\ref{OP1}) is a corollary of Theorem 1.

\begin{corollary}
The optimal allocations for the BS transmit power, $p_0^*(i)$, the duration of the EH phase, $\tau_0^*(i)$, and the duration of the IT phases, $\tau_k^*(i)$, are given by Theorem 1, where $\bar R_1 = \bar R_2 = \cdots = \bar R_K = 1$.
\end{corollary}
\begin{IEEEproof}
Similar to the proof of Theorem 1, we firstly assume that constraint $C1$ is strictly positive, which is validated by the optimal solution. Introducing the change of variables $e(i) = p_0(i)\tau_i(i)$, (\ref{OP1}) is transformed into a convex optimization problem. Its Lagrangian corresponds to (\ref{lagrang}) with the $\log$ operator omitted (yielding sum-rate maximization, instead of sum-log-rate maximization). As a result, the Lagrangian derivatives with respect to $e(i)$, $\tau_k(i)$, and $\tau_0(i)$ correspond to (\ref{ravenka_e_i}), (\ref{ravenka_tau_ki}), and (\ref{ravenka_tau_0i}), respectively, with $\bar R_k$ set to unity.
\end{IEEEproof}

The practical implementation of Corollary 1 is similar to Algorithm 1, with $\bar R_k = 1, \forall k$ (and step 12 removed).

Note that, when $p_c=0$, (\ref{OP1}) reduces to [\ref{lit1a}, Eq. (3)], and its solution is given by [\ref{lit1a}, Theorem 1]. Additionally, the solution of (\ref{OP1}) in the special case of $K=1$ is given by [\ref{lit1a}, Theorem 2], but a similar proof is not applicable to the case of arbitrary $K$. Corollary 1 presents the general solution of the sum rate maximization problem for arbitrary values of $p_c$ and $K$.

\vspace{-2mm}
\section{Numerical Results}
For the numerical examples, we assume a Rayleigh fading environment, with the path-loss exponent $\alpha =3$ and the path loss of 30dB at reference distance of $1$m, i.e., $E[x_k'(i)] =10^{-3}\, D_k^{-\alpha}$. Five EHUs are placed at different distances from the BS: $D_1=10$m, $D_2 = 12.5$m, $D_3=15$m, $D_4=17$m, and $D_5=18.8$m. We also set $P_{max} = 5 P_{avg}$, and $N_0 = 10^{-12}$W. As the measure for the system fairness, we adopt the Jain's fairness index [\ref{nov1}], $J = (\sum_{k=1}^K \bar R_k)^2 / (K \sum_{k=1}^K \bar R_k^2)$.

Fig. 1 considers the effect of the processing cost at $P_{avg} = 1$W. Fig. 1a depicts the sum rate over the uplink of the WPCN, $(1/M) \sum_{k = 1}^K \sum_{i = 1}^M r_k(i)$, whereas Fig. 1b depicts the system fairness, $J$. As $p_c$ increases, the sum rates steadily decrease, while the fairness index is kept nearly constant for both proposed protocols. For a given $K$, the sum rate difference and the fairness index difference between the two proposed protocols are nearly independent of $p_c$ (i.e., the all the curves are nearly parallel). As $K$ increases, the {\it max-sum-rate} protocol exerts its increasing advantage over the {\it PF} protocol in terms of the sum rate. On the other hand, relative to the {\it max-sum-rate} protocol, the {\it PF} protocol attains a higher level of system fairness for all $K$, and this advantage increases with $K$.

Fig. 2 depicts the uplink sum rate vs. $P_{avg}$ for various $p_c$ as parameter, by assuming that all EHUs are at the same distance from BS (i.e., $D_k=10$m, $\forall k$, and, therefore, $J = 1$). The two proposed protocols are compared against the benchmark protocol studied in [\ref{lit2}], which maximizes the uplink sum rate by fixing the output power at the BS (i.e., $p_0(i) = P_0 = const., \forall i$), and adjusting only the durations of the EH/IT phases, $\tau_k(i)$, as per [\ref{lit2}, Eq. (10)]. Note, the benchmark only applies to the case of $p_c = 0$, but not $p_c > 0$. For a fair comparison, $P_0$ is selected in order to satisfy our average power constraint $C2$, i.e., $E[P_0  \tau_0(i)] = P_{avg}$. Fig. 2 indicates that the advantage of the {\it max-sum-rate} protocol in terms of the sum rate is less pronounced with increasing $P_{avg}$, especially for larger values of $K$.

\begin{figure}
\vspace{0mm}
\begin{minipage}[b]{.45\linewidth}
\centering
\includegraphics[trim = 20mm 10mm 35mm 50mm, scale=0.35]{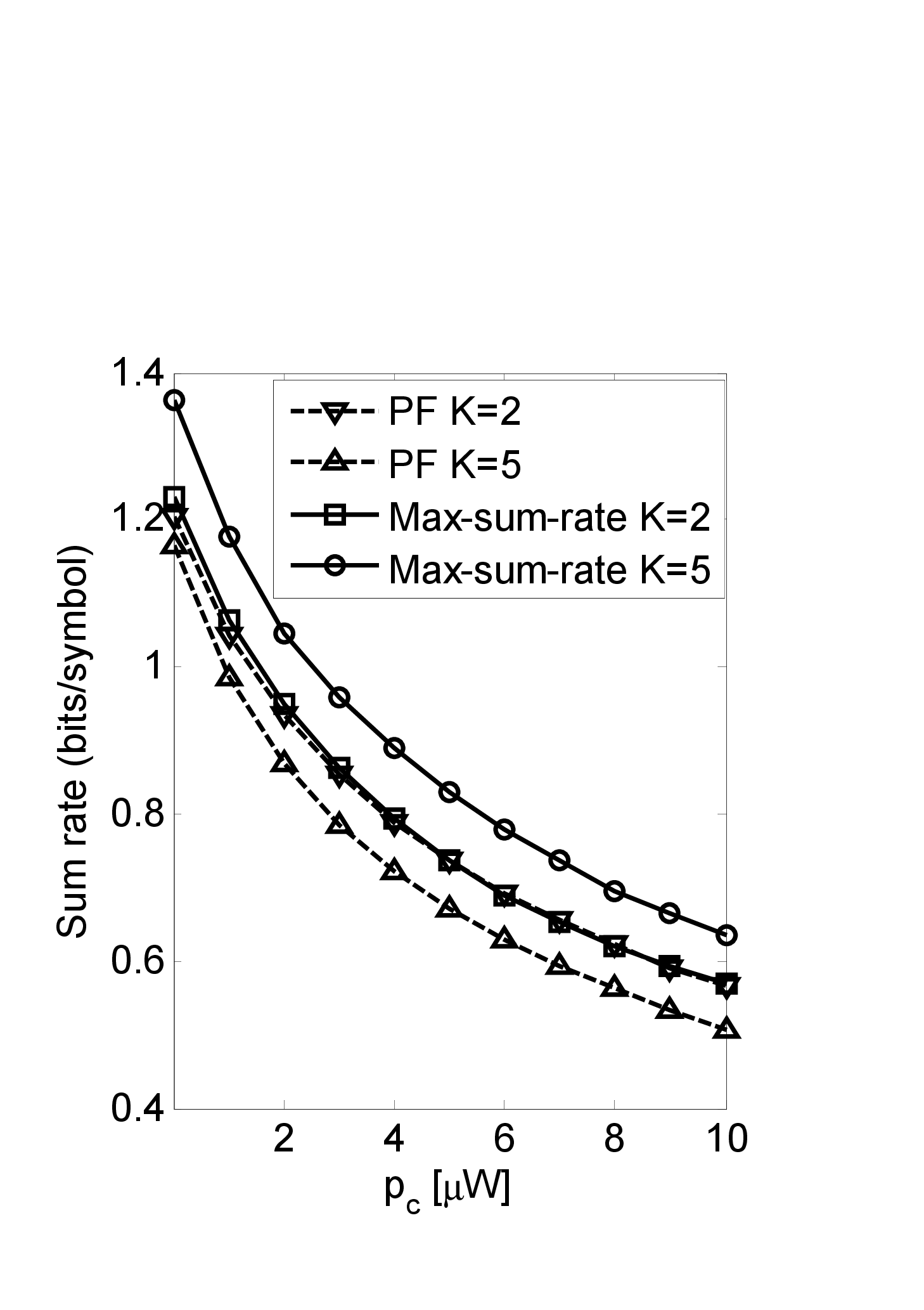}
\text{(a) Sum rate}
\end{minipage}
\begin{minipage}[b]{.45\linewidth}
\centering
\includegraphics[trim = 15mm 10mm 40mm 50mm, scale=0.35]{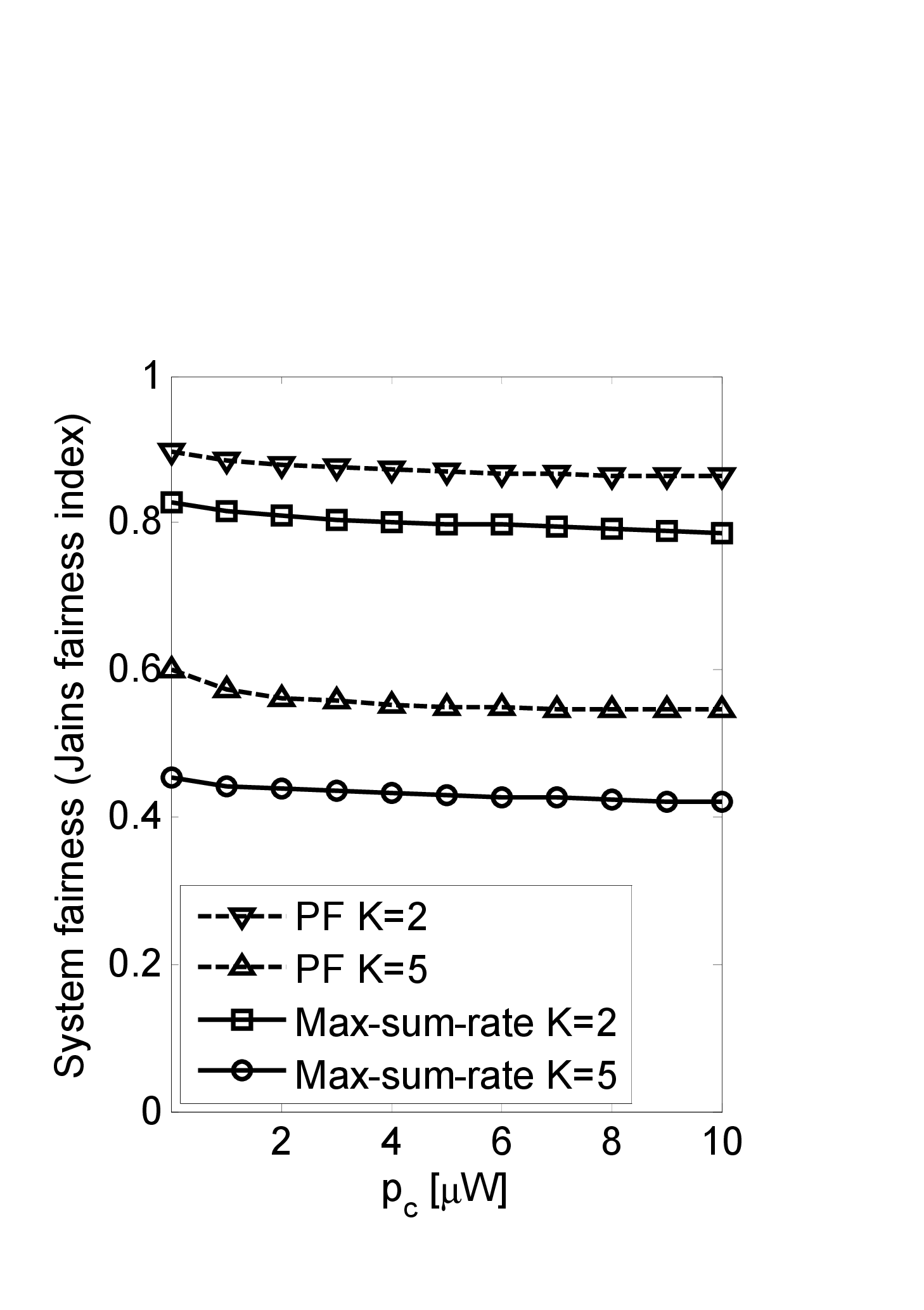}
\text{(b) System fairness}
\end{minipage}
\caption{Tradeoff between the sum rate and fairness } \label{fig2}
\end{figure}

\begin{figure}[tbp]
\centering
\includegraphics[scale=0.47]{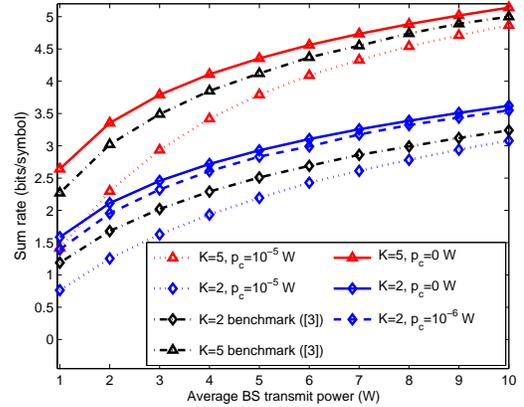}
\caption{The effect of BS average transmit power on sum rate} \vspace{-4mm}
\label{fig1}
\end{figure}

\vspace{-3mm}
\section{Conclusion}
\vspace{-1mm} In this paper, we have proposed two WPCN protocols that guarantee either proportionally fair resource allocation among EHUs at different distances from the BS, or sum rate maximization over the uplink. The optimal power and time allocations of both protocols account for the non-ideal circuit power consumption at the EHUs, which facilitates the practicality of these protocols.

\vspace{-3mm}
\appendices
\section{Proof of Theorem 1}
\label{sec_A}
\vspace{-0mm}
Let us assume the optimal solution of (\ref{OP2}) is such that the constraint $C1$ is strictly positive, i.e.,
\begin{equation} \label{Pk_assump}
P_{k}(i) =\frac{\eta_k x_{k}(i)N_0 p_{0}(i)\tau_{0}(i)}{\tau_{k}(i)}-p_c > 0, \quad  \forall i, \forall k.
\end{equation}
In this case, after applying (\ref{Pk_assump}) and the variables change $e(i) = p_{0}(i)\tau_{0}(i)$, (\ref{OP2}) is transformed into

\begin{eqnarray} \label{OP3}\notag
\underset{\tau_{k}(i),\tau_{0}(i),e(i)} {\text{Maximize}} \ \hspace{-3mm} & & \hspace{-3mm}\sum_{k=1}^K\log\left(\frac{1}{M}\sum_{i=1}^M\tau_{k}(i)\right.\\\notag
& &\left. \times \log_2 \left(1-x_{k}(i)p_c+\frac{a_{k}(i)e(i)}{\tau_{k}(i)} \right)\right) \notag
\end{eqnarray}
\text{s.t.}
\vspace{-5mm}
\begin{eqnarray}
\begin{array}{ll}
&\bar{C}2: \frac{1}{M}\sum_{i=1}^M e(i) \leq P_{avg} \\ \notag
&\bar{C}3: 0 \leq e(i) \leq P_{max} \, \tau_{0}(i), \forall i \\ \notag
&\bar{C}4: \sum_{k=1}^K \tau_{k}(i) = 1 -\tau_{0}(i), \forall i \\ \notag
&\bar{C}5: 0 < \tau_{n}(i) < 1, \forall i, \, 0 \leq n \leq K.
\end{array}
\end{eqnarray}
\vspace{-7mm}
\begin{equation}
\end{equation}
\hspace{3mm}
The function $\tau_k(i) \log_2(1 - p_c x_k(i) + a_k(i) e(i)/\tau_k(i))$ is the perspective of $\log_2(1 - p_c x_k(i) + a_k(i) e(i))$, and, therefore, it is jointly concave in $e(i)$ and $\tau_k(i)$ [\ref{lit3a}, Section 3.2.6]. The inner sum of the objective function of (\ref{OP3}) is positive and concave in $e(i)$ and $\{\tau_k(i)\}_{k = 1}^K$, and, therefore, its logarithm is also concave [\ref{lit3a}, Section 3.5.1]. The constraints are all affine (i.e., convex) functions. As a result, (\ref{OP3}) is a convex optimization problem, which can be solved by applying the Lagrangian dual method. Its Lagrangian is given by
\begin{eqnarray} \label{lagrang} \notag
\mathcal{L} \hspace{-3mm}  &=& \hspace{-3mm}  \sum_{k=1}^K\log\left(\frac{1}{M}\sum_{i=1}^M\tau_{k}(i)\log_2 \left( 1-x_{k}(i)p_c+\frac{a_{k}(i)e(i)}{\tau_{k}(i)} \right) \right)\\ \notag
&-& \hspace{-3mm} \lambda \left(\frac{1}{M}\sum_{i=1}^Me(i)-P_{avg}\right) - \sum_{i=1}^M\beta_{i} (e(i)-P_{max}\tau_{0}(i)) \\
&+& \hspace{-3mm} \sum_{i=1}^M\alpha_{i}e(i) - \sum_{i=1}^M \mu_i \left(\tau_{0}(i)+\sum_{k=1}^K\tau_{k}(i)-1\right)
\end{eqnarray}
where the Lagrange multipliers $\lambda$, $\alpha_i$, $\beta_i$, and $\mu_i$ are associated with $\bar{C}2$, the left hand side of $\bar{C}3$, the right hand side of $\bar{C}3$, and $\bar{C}4$, respectively. They satisfy the following slackness conditions: $\lambda (\sum_{i=1}^M e(i)/M -P_{avg}) = \alpha_{i}e(i) = \beta_{i} (e(i)-P_{max}\tau_{0}(i))= \mu_i (\tau_{0}(i) + \sum_{k=1}^K \tau_{k}(i) - 1 ) = 0$. The derivatives of (\ref{lagrang}) with respect to $e(i)$, $\tau_k(i)$, and $\tau_0(i)$, set equal to zero, yield the following equation set:
\begin{equation}
\label{ravenka_e_i}
\sum_{k=1}^K\frac{1}{\bar R_k}\frac{a_{k}(i)}{1-p_cx_{k}(i)+\frac{a_{k}(i)e(i)}{\tau_{k}(i)}}-\lambda+\alpha_i-\beta_i = 0
\end{equation}
\begin{equation}
\label{ravenka_tau_ki}
\hspace{-5mm} \log\left(1-x_{k}(i)p_c+\frac{a_{k}(i)e(i)}{\tau_{k}(i)}\right) -  \frac{\frac{a_{k}(i)e(i)}{\tau_{k}(i)}}{1-x_{k}(i)p_c+\frac{a_{k}(i)e(i)}{\tau_{k}(i)}} = \bar R_k \mu_i
\end{equation}
\begin{equation}
\label{ravenka_tau_0i}
\beta_iP_{max} - \mu_i = 0.
\end{equation}

\underline{Case 1}: Let $e(i) = 0$. Then, no power is allocated to epoch $i$, yielding $p_0^*(i)=0$.

\underline{Case 2}: Let $e(i) = P_{max} \tau_{0}(i)$. Due to the slackness conditions, we obtain $\alpha_i = 0$, $\beta_i > 0$, and $\mu_i > 0$. Thus, (\ref{ravenka_tau_ki}) is reduces to
\begin{equation} \label{transc1}
\log(1-x_{k}(i)p_c+z_{k}(i)) -\frac{z_k(i)}{1-x_{k}(i)p_c+z_{k}(i)} = \beta_iP_{max}\bar R_k,
\end{equation}
where $z_{k}(i) = a_{k}(i)P_{max}\tau_{0}(i)/\tau_{k}(i)$. The closed form solution of (\ref{transc1}) is given by (\ref{sol41}). From (\ref{ravenka_e_i}), we obtain
\begin{equation}
\label{rav_uslov}
\beta_i=\sum_{k=1}^K\frac{1}{\bar R_k}\frac{a_{k}(i)}{1-p_cx_{k}(i)+z_k(i)}-\lambda > 0.
\end{equation}

\noindent The combination of (\ref{rav_uslov}) and (\ref{sol41}) yields (\ref{sol51}). Additionally, (\ref{rav_uslov}) leads to the condition for $p_0(i) = P_{max}$ in (\ref{sol11}).

In order to verify the assumption (\ref{Pk_assump}), we notice that, since $\beta_i > 0$, the left hand side of (\ref{transc1}) should be positive, i.e.,
\begin{equation} \label{transc2}
\log(1 - x_{k}(i)p_c + z_{k}(i)) -\frac{z_k(i)}{1 - x_{k}(i)p_c + z_{k}(i)} > 0.
\end{equation}
For arbitrary $x_{k}(i)p_c > 0$, (\ref{transc2}) is satisfied iff $z_{k}(i) - x_{k}(i)p_c > 0$, which is equivalent to the assumption (\ref{Pk_assump}).

\underline{Case 3}: Let $0< p_0 (i) < P_{max}$. Due to the slackness conditions, $\alpha_i = \beta_i = 0$. In this case, (\ref{ravenka_e_i}) and (\ref{ravenka_tau_ki}) can be satisfied only for one specific set of values for $\{x_{k}(i)\}_{k = 1}^K$. Since the channel gains are assumed to follow continuous PDFs, occurrence probability of this case is thus zero.

\end{document}